\documentclass[aps,preprint]{revtex4}
\usepackage{graphicx}
\usepackage{epsfig}

\begin{document}

\title{Kerr-Newman-de Sitter black holes in $ f(R) $ gravity with constant curvature: horizon  structure and extremality}
\author{\large Alikram N. Aliev}
\address{Department of Basic Sciences, Faculty of Engineering and Natural Sciences,
Maltepe University, 34857 Maltepe, Istanbul, T\"{u}rkiye}
\author{\large  G\"{o}ksel Daylan Esmer}
\address{Department of Physics, Istanbul University, Vezneciler, 34134 Istanbul,T\"{u}rkiye}

\date{\today}

\begin{abstract}

The theory of  $ f(R) $ gravity with constant curvature (i.e.\ constant scalar curvature) admits rotating and charged black hole solutions obtained from the Kerr-Newman-(A)dS metrics of general relativity through appropriate rescalings of the metric parameters. In this paper, we focus on the Kerr-Newman-de Sitter case and present a unified analytic treatment of the horizon structure and its physical properties, allowing for a transparent comparison between general relativity and  $ f(R)$ gravity with constant curvature. We solve the quartic equation determining the horizon locations and derive closed analytic expressions for the horizon radii. Focusing on extremal configurations, we obtain analytic formulas for the squared rotation parameter $ a^2 $  and the inverse square of the curvature radius  $ l^{-2} $  as functions of the horizon location and the electric charge. For generic values of these parameters, the extremality conditions are non-universal, reducing to the familiar Kerr-Newman bound only in the limit of vanishing background curvature. We identify an ultra-extremal configuration in which  $ a^2 $ attains its maximal value at zero charge and decreases monotonically to zero as the charge approaches its limiting value, while $ l^{-2 }$ increases correspondingly.  As an illustrative example, we show that black holes with charge $ q=M/2 $ necessarily possess a minimum rotation, which emerges naturally as an intersection point in our analytic description of $ a^2 $  and $ l^{-2 }$, when embedded in a universe characterized by a critical value of $ l^{-2} $ (equivalently, the scalar curvature or the cosmological constant).  Finally, we demonstrate that when the mass satisfies $ M^2= (a^2+q^2)(1-a^2/l^2)$, the quartic horizon equation factorizes, leading in the extremal regime to a chiral-like horizon structure that allows only the outer-cosmological horizon merger.

\end{abstract}

\maketitle
\newpage

\section{Introduction}

Black holes are among the most profound predictions of general relativity (GR) and are well described, in the asymptotically flat case, by the Kerr–Newman family of exact solutions to the Einstein field equations. These solutions possess remarkable uniqueness and stability properties and admit hidden symmetries, which enable the complete separation of variables in the geodesic equations \cite{israel, hawking, carter} (see also Ref. \cite{fn}). As a result, the exterior spacetimes of black holes are widely regarded as mathematically elegant and well understood. By contrast, the near-zone dynamics and the interior structure of black holes remain uncertain, as classical GR is expected to break down in strong-gravity regimes. Moreover, GR is incompatible with quantum mechanics and is therefore expected to require modifications at both ultraviolet and infrared scales \cite{stelle, fn}.

All this provides strong motivation to further scrutinize black hole spacetimes in the strong-gravity regime, both within GR and beyond it. Fortunately, significant progress has been achieved in this direction over recent years. The direct detection of gravitational waves from merging black holes in binary systems by the Laser Interferometer Gravitational-Wave Observatory (LIGO) \cite{abbott1}, together with the first horizon-scale images of the supermassive black holes at the centers of the M87 galaxy and the Milky Way obtained by the Event Horizon Telescope (EHT) collaboration \cite{akiyama1}, have opened a new era of black hole exploration (see  Refs. \cite{afshordi, sunny} for recent reviews).

These observational advances have also dramatically revitalized efforts to test the fundamental paradigms of black holes, as well as possible deviations from the predictions of GR. In this context, considerable attention has been devoted to exploring black hole solutions in alternative theories of gravity. For example, in certain scalar-tensor theories with a minimally coupled scalar field, black hole solutions remain indistinguishable from their GR counterparts at the level of stationary spacetimes; nevertheless, the additional scalar degree of freedom manifests itself dynamically through phenomena such as scalarization in the inspiral-merger-ringdown waveforms  of neutron star or black hole binaries \cite{lehner1}. In other classes of modified gravity theories, such as Einstein-Chern-Simons and Einstein-Gauss-Bonnet gravity, the coupling between gravity and a scalar field can trigger dynamical scalarization of rotating black holes under specific conditions \cite{stein, silva}. Numerical simulations of binary mergers involving such black holes reveal clear imprints of the scalar degree of freedom, most notably through dipole radiation, which is entirely absent in GR \cite{yunes, pompili}. Similar analyses of deviations from general relativity in the inspiral and merger dynamics of compact binaries have also been carried out within other classes of higher-derivative gravity theories. Among these, $ f(R) $ gravity is of particular interest, as it represents one of the simplest and most consistent extensions of GR. Notably, the theory admits a well-posed initial value formulation, rendering it suitable for numerical simulations of fully nonlinear dynamics. This has enabled, for instance, detailed studies of neutron star mergers within $ f(R) $ gravity \cite{lehner2}.

Motivated by these developments in precision black hole science, we examine rotating and charged black holes in $ f(R) $ gravity, with particular emphasis on the structure of their horizons and the associated physical properties. The action of this theory is obtained by replacing the scalar curvature in the Einstein-Hilbert action with $ R + f(R) $, where $ f(R) $ is an arbitrary function of $ R $. The resulting field equations contain higher-order derivatives of the metric, typically up to fourth order, which makes the construction of exact analytic black hole solutions a highly nontrivial task (see, for instance,~\cite{nojiri}). Remarkably, a variety of intriguing static and spherically symmetric solutions, with or without constant curvature, including the Schwarzschild and Reissner–Nordström black holes, have nevertheless been obtained in a number of works \cite{maroto, zerbini, odin1, tang}. For stationary and axially symmetric spacetimes, a particular class of solutions with constant curvature can likewise be found, corresponding to the Kerr and Kerr–Newman black holes. In this case, the constant scalar curvature $ R_0 $ can be expressed in terms of an effective cosmological constant, determined by the function $ f(R) $ and its first derivative evaluated at $ R=R_0 $. This situation remains unchanged when electric charge is included, since the trace of the corresponding energy-momentum tensor vanishes identically in four dimensions. Consequently, after appropriate rescalings of the metric parameters, the Kerr-(A)dS and Kerr-Newman-(A)dS metrics of general relativity also provide solutions of $ f(R) $ gravity \cite{larran, cembra}.

In this paper, we focus on the Kerr-Newman-de Sitter case and present a unified analytic treatment of the horizon structure and its physical properties, allowing for a transparent comparison between general relativity and  $ f(R)$ gravity with constant curvature. The paper is organized as follows. In Sec. II, we briefly review the field equations of $ f(R) $ gravity coupled to a Maxwell field and show that, for constant scalar curvature $ R=R_0 $, they reduce to the Einstein field equations with an effective cosmological constant and a suitably rescaled Newton constant (or, equivalently, energy-momentum tensor). This demonstrates that constant curvature solutions of GR, including the Kerr-Newman-(A)dS family, can be mapped into $ f(R) $ gravity through appropriate rescalings of the metric parameters. In this section, we also introduce the inverse square of the curvature radius $ l^{-2} $, which is proportional to the scalar curvature (or, equivalently, the cosmological constant), providing a unified framework for treating both Kerr-Newman-de Sitter black holes in GR and their counterparts in $ f(R) $ gravity. In Sec. III, we solve the quartic equation determining the horizon locations and derive closed analytic expressions for the horizon radii. By expanding these expressions in powers of the curvature radius $ l $, we clarify their physical interpretation in terms of the inner and outer black hole horizons and the cosmological horizon. We then focus on extremal configurations and obtain analytic formulas for the squared rotation parameter $ a^2 $ and the inverse square of the curvature radius $ l^{-2}$ as functions of the horizon location and the electric charge. These results show that the extremality condition for $ a^2 $ becomes universal only in the limit of vanishing background curvature, reproducing the familiar Kerr-Newman bound of GR. We further identify an ultra-extremal configuration in which $ a^2 $ attains its maximal value at vanishing electric charge and decreases monotonically to zero as the charge approaches its limiting value, while $ l^{-2} $ increases correspondingly. As an illustrative example, we show that black holes with charge $ q=M/2 $ necessarily possess a minimum rotation when embedded in a universe characterized by a critical value of $l^{-2}$ (equivalently, the scalar curvature or the cosmological constant). A comparison with the uncharged case reveals that the presence of electric charge increases the minimum rotation required for the black hole. Finally, we demonstrate that when the mass satisfies $ M^2=(a^2+q^2)(1-a^2/l^2) $, the quartic horizon equation factorizes into two independent pairs of roots. In the extremal regime, this leads to a chiral-like structure in the horizon configuration space, allowing only the outer–cosmological horizon merger while excluding the inner-outer one. Our discussion and conclusions are presented in Sec. IV.

\section{Field equations and the spacetime metric}

We begin with the action for $ f(R) $ gravity coupled to a Maxwell field,
\begin{eqnarray}
S &=&  \int d^4 x \sqrt{-g} \left[\frac{1}{16 \pi G}\,\left (R + f(R) - 2 \Lambda \right) 
- \frac{1}{16 \pi} \,  F_{\mu\nu} F^{\mu\nu} \right] ,
\label{action}
\end{eqnarray}
where $ R = g^{\mu\nu} R_{\mu\nu} $ is the scalar curvature, $ f(R) $ is an arbitrary function of $ R $
subject to viability conditions ensuring consistent gravitational and cosmological dynamics \cite{star},
$ F_{\mu\nu} = 2 \partial_{[\mu} A_{\nu]} $ is the electromagnetic field tensor, and $ \Lambda $ denotes the cosmological constant.

The field equations are obtained by varying this action with respect to the metric. We have
\begin{eqnarray}
\label{eq1}
  && R_{\mu\nu} \left( 1 + f'(R) \right) - \frac{1}{2}\,  g_{\mu\nu} \left( R+f(R) - 2 \Lambda \right)  + \left( g_{\mu\nu} \, 
 \nabla^2 - \nabla_\mu \nabla_{\nu} \right) f'(R)    =  8\pi G T_{\mu\nu}\, ,  \\ [3mm] 
&& \nabla_{\nu} F^{\mu \nu}  =  0 \,, 
 \label{eq2}
\end{eqnarray}
where $  f'(R)  $ denotes the differentiation of  $ f(R) $  with respect to $ R $,  $ \nabla_{\mu} $ stands for  a covariant derivative operator,  $ \nabla^{2}= g^{\mu\nu}\nabla_{\mu} \nabla_{\nu}$,  and the electromagnetic energy-momentum tensor is given by
\begin{eqnarray}
T_{\mu \nu} &=&   \frac{1}{4\pi}\left(F_{\mu \lambda}F_{\nu}^{~\lambda} - \frac{1}{4} \,g_{\mu \nu} \, F_{\alpha \beta} F^{\alpha\beta}\right) .
\label{emt}
\end{eqnarray}
The trace of the energy--momentum tensor $T_{\mu\nu}$ vanishes identically. 
Taking the trace of Eq. (\ref{eq1}), we obtain
\begin{eqnarray}
3\, \nabla^{2} f'(R) + R \left(1 + f'(R) \right)
&=& 2 \left( R + f(R) - 2\Lambda \right),
\label{treq1}
\end{eqnarray}
which, for constant scalar curvature $ R = R_0 $, reduces to
\begin{eqnarray}
R_0 &=& \frac{2\left[f(R_0) - 2\Lambda \right]}{f'(R_0) - 1} \,.
\label{constantR}
\end{eqnarray}
Using this expression, it is straightforward to show that for  constant scalar curvature $ R=R_0 $,  the field equations in (\ref{eq1}) can be transformed into the form 
\begin{eqnarray}
 R_{\mu \nu}& = &\frac{1}{2}\,  g_{\mu \nu}\, \frac{f(R_0) - 2\Lambda}{f'(R_0)  - 1} +  \frac{8\pi G  T_{\mu \nu}}{1+f'(R_0)}\,.
\label{einform}
\end{eqnarray}
Next, we introduce an ``effective'' cosmological constant $ \lambda $ defined by
\begin{eqnarray}
\lambda &=& \frac{1}{2}\, \frac{f(R_0) - 2\Lambda}{f'(R_0) - 1}
= \frac{R_0}{4} \,,
\label{effcosmcon}
\end{eqnarray}
where we have also used Eq. (\ref{constantR}). It follows that $ \lambda $ does not vanish even when the bare cosmological constant $ \Lambda $ is set to zero. In this case, the effective cosmological constant is entirely induced by the $ f(R) $ modification of gravity. It is now worth noting that, together with Eq. (\ref{effcosmcon}) and an appropriate rescaling of the Newtonian constant $ G $ (or, equivalently, of the energy--momentum tensor $ T_{\mu\nu} $) to absorb the constant factor $ 1 + f'(R_0) $, the field equations in Eq. (\ref{einform}) can be cast into the standard Einstein form of general relativity. Thus, we conclude that the constant curvature solutions of general relativity, including the Kerr-Newman-(A)dS family of black holes, can be mapped into $ f(R) $ gravity through an appropriate rescaling of the metric parameters \cite{larran, cembra}. 
With this in mind, we focus exclusively on the Kerr--Newman--de Sitter case.
The corresponding spacetime metric in $f(R)$ gravity with constant curvature,
written in Boyer--Lindquist coordinates, takes the form
\begin{eqnarray}
ds^2 & = & -{{\Delta_r}\over {\Sigma}} \left(\,dt - \frac{a
\sin^2\theta}{\Xi}\,d\phi\,\right)^2 + {\Sigma \over~ \Delta_r}
dr^2 + {\Sigma \over ~\Delta_{\theta}}\,d\theta^{\,2} +
\frac{\Delta_{\theta}\sin^2\theta}{\Sigma} \left(a\, dt -
\frac{r^2+a^2}{\Xi} \,d\phi \right)^2 \,,\nonumber\\
 \label{4knads}
\end{eqnarray}
where
\begin{eqnarray}
\Delta_r &= &\left(r^2 + a^2\right)\left(1 -  \frac{ r^2}{l^2}\right)
- 2 M r  +\frac{Q^2} { 1+f'(R_0) }\,,~~~~~\Sigma = r^2+ a^2 \,\cos^2\theta \,,
\nonumber \\[2mm]
\Delta_\theta & = & 1 +  \frac{ a^2}{l^2}
\,\cos^2\theta\,,~~~~~ \Xi=1 +
\frac{ a^2}{l^2} \,. 
\label{metfunct}
\end{eqnarray}
Meanwhile, the associated electromagnetic field is described by the potential one-form
\begin{equation}
A = -\,\frac{Q\, r}{\Sigma}
\left( dt - \frac{a \sin^{2}\theta}{\Xi}\, d\phi \right) .
\label{potform}
\end{equation}
The parameters $M$, $a$, and $Q$ appearing in the above expressions are related to the physical mass, angular momentum, and electric charge of the black hole, respectively \cite{aliev1}. For convenience, we introduce the curvature radius $l$, associated with the
de Sitter background curvature.  For positive scalar curvature $R_0>0$, or equivalently a positive cosmological constant $\Lambda>0$,  it is defined as
\begin{eqnarray}
l^{-2} = \frac{R_0}{12} = \frac{\Lambda}{3}\,,
\label{l}
\end{eqnarray}
where we have used $R_0 = 4\Lambda$, as given in Eq. (\ref{effcosmcon}).  With this definition of the inverse square curvature radius, we obtain a unified framework for treating the Kerr-Newman-de Sitter family of black holes in general relativity and their counterparts in $f(R)$ gravity. In addition, as follows from the expression for $\Delta_r$ given in Eq.~(\ref{metfunct}), the charge parameter appearing in the metric is multiplied by the factor $(1+f'(R_0))^{-1/2}$. This arises from a suitable rescaling of the energy--momentum tensor $T_{\mu\nu}$ (rather than the Newton constant $G$, which is set to unity) in order to absorb this factor in Eq. (\ref{einform}).   We note that the factor $\Xi$ appears in the metric to eliminate conical singularities along the axis of symmetry.  Finally, throughout this work we restrict our analysis to the Kerr-Newman-de Sitter geometry, corresponding to spacetimes with positive scalar curvature or, equivalently, a positive cosmological constant.

Before concluding this section, it is worth commenting on the relation between black hole solutions in $f(R)$ gravity with constant curvature and black holes embedded in dark-energy-motivated backgrounds. In the latter case, dark energy is often modeled phenomenologically by an effective or imperfect fluid added to the Einstein equations, leading to modified Reissner-Nordström-like or Kerr-like  geometries (see, for instance,  \cite{kisel, bobo}). While such approaches provide valuable insight into the possible influence of dark energy on black hole spacetimes, the underlying physical interpretation differs substantially from that in modified gravity theories \cite{nojiri}. In the present work, the effective cosmological constant arises purely from the gravitational sector through the constant-curvature condition, without introducing additional matter degrees of freedom beyond the electromagnetic field. As a result, although the horizon structure may exhibit qualitative similarities, the solutions studied here represent exact vacuum (up to electromagnetism) geometries of $f(R)$ gravity rather than black holes supported by prescribed dark-energy matter sources.

\section{Horizons and Extremality}

One of the challenging aspects of black holes in GR is the existence of a lower bound on the rotation parameter, such that over-rotating solutions are excluded as they lead to naked singularities. For instance, the Kerr--Newman solution satisfies the bound $a^2 \leq M^2 + Q^2$, where the equality $a^2 = M^2 + Q^2$, or simply $a^2 = M^2$ in the absence of electric charge ($Q=0$), corresponds to maximally rotating (extremal) black holes with vanishing temperature.  On the other hand, astronomical observations strongly suggest that nearly maximally rotating black holes are common, both in X-ray binaries and at the centers of most galaxies \cite{jef, daly}. Remarkably, in some cases the inferred spin values approach the Kerr bound of GR, raising the question of possible astrophysical violations of this limit and motivating the study of black holes beyond GR \cite{gimon, an2}. It is particularly intriguing that a violation of the Kerr bound can already occur for Kerr-de Sitter black holes within GR itself \cite{akcay, slany}.  With this motivation, in the following we present a unified analytic treatment of the horizon structure and extremal configurations of Kerr-Newman-de Sitter black holes, both in GR and in $f(R)$ gravity with constant curvature.

\subsection{Horizons I}

As follows from Eqs.~(\ref{4knads}) and (\ref{metfunct}), the horizon locations are determined by the real roots of the quartic equation $\Delta_r=0$, which can be written in the form
\begin{eqnarray}
r^4 + a_2\, r^2 + a_1\, r + a_0 &=& 0 \, ,
\label{quartic1}
\end{eqnarray}
where the constant coefficients are given by
\begin{eqnarray}
a_2 &=& a^2 - l^2 \,, \qquad
a_1 = 2 M l^2 \,, \qquad
a_0 = -\left(a^2 + q^2\right) l^2 \, .
\label{constcoef}
\end{eqnarray}
For convenience, we have also introduced the charge parameter  $ q $, defined as 
\begin{eqnarray}
q^2 = \frac{Q^2}{ 1+f'(R_0)}\,.
\label{chargepara}
\end{eqnarray}
In the general relativistic limit $f'(R_0)=0$, the definition (\ref{chargepara}) reduces to the standard Kerr–Newman electric charge. The quartic equation in Eq. (\ref{quartic1}) generally admits four roots, which satisfy the Vieta relations. Using the coefficients given in Eq. (\ref{constcoef}), we obtain
\begin{eqnarray}
&& r_1 + r_2 + r_3 + r_4 = 0 \,, \qquad
r_1 r_2 r_3 r_4 = -\left(a^2 + q^2\right) l^2 \,, \nonumber \\[2mm]
&& \sum_{i<j} r_i r_j = a^2 - l^2 \,, \qquad
\sum_{i<j<k} r_i r_j r_k = -2 M l^2 \, .
\label{vieta1}
\end{eqnarray}
From the relations in the first line of Eq. (\ref{vieta1}), it follows that the quartic equation (\ref{quartic1}) necessarily possesses at least three real positive roots and one real negative root. We are interested in determining these roots explicitly, which, using Cardano's method, can be expressed in terms of a real root of the associated resolvent cubic equation (see, e.g., Ref. \cite{astegun}).
\begin{equation}
u^3 - a_2  u^2 -  4 a_0  u  - (a_1^2 - 4 a_0  a^2)  =  0 .
\label{cubic1}
\end{equation}
The vanishing of the discriminant of this equation allows us to introduce two extremal mass parameters, given by
\begin{eqnarray}
M_{1e}^2 &= & \frac{l^2}{54}\left(\zeta - \eta^3 \right) \,,~~~~~
M_{2e}^2 = \frac{l^2}{54}\left(\zeta + \eta^3 \right)     \,,
\label{exmasses}
\end{eqnarray}
where
\begin{equation}
\label{zeta} 
\zeta = \left(1- \frac{a^2}{l^2} \right)\left[ \left(1- \frac{a^2} {l^2} \right)^2 +\frac{36  \left (a^2+ q^2 \right)}{l^2}\right],~~~
\eta =\left[\left(1 - \frac{a^2}{l^2} \right)^2 - \frac{12 \left(a^2+q^2\right)}{l^2}\right]^{1/2}.
\label{eta}
\end{equation}
It is straightforward to show that $ M_{1e}^2 \to a^2 + q^2 $ in the limit $ l \to \infty $, whereas $ M_{2e}^2 $ diverges as $ M_{2e}^2 \to l^2/27 $ in this limit. In terms of these quantities, the real root of the associated resolvent cubic equation (\ref{cubic1}) can be written in the following concise form
\begin{eqnarray}
u &=&  \frac{ a^2 -l^2}{3}+\frac{l^{ 4/3} \left(M_{2e}^2 - M_{1e}^2
\right)^{2/3}}{\left(2 N^2- M_{1e}^2-M_{2e}^2 \right)^{1/3}}\, + \,
l^{4/3} \left(2 N^2- M_{1e}^2-M_{2e}^2 \right)^{1/3} ,
\label{resolvereal}
\end{eqnarray}
where
\begin{equation}
N^2=M^2+\sqrt{\left(M^2-M_{1e}^2\right)\left(M^2-M_{2e}^2\right)}\,\,.
\label{censor1}
\end{equation}
We note that $ M_{1e}^2 < M_{2e}^2 $, and that cosmic censorship requires the black
hole mass parameter to satisfy
\begin{equation}
 M_{1e} \leq M \leq M_{2e} \,,
 \label{censor2}
\end{equation}
where equality corresponds to the extremal configurations. To proceed further,
it is convenient to introduce the following notations
\begin{eqnarray}
X &=& \sqrt{u + l^2 - a^2}\,, \nonumber \\[3mm]
Y &=& \sqrt{-u + l^2 - a^2 + \frac{4 M l^2}{X}}\,, \nonumber  \\[3mm]
Z &=& \sqrt{-u + l^2 - a^2 - \frac{4 M l^2}{X}}\,.
\label{4rootsnot}
\end{eqnarray}
These quantities allow us to express the four real roots of the quartic equation
(\ref{quartic1}) in the compact form
\begin{eqnarray}
r_1&=&-\frac{1}{2}\left(X+Y\right),  \qquad  r_3= \frac{1}{2}\left(X-Z \right),
 \nonumber \\[2mm]
r_2 & = & -\frac{1}{2}\left(X-Y
\right),   \qquad r_4= \frac{1}{2}\left(X+Z \right), 
\label{4roots}
\end{eqnarray}
To clarify the physical interpretation of these roots in terms of the inner and outer black hole horizons and the cosmological horizon, we expand them in powers of the curvature radius in the limit $ l \to \infty $.  After some algebra  and retaining terms up to second order in $ l $, we find
\begin{eqnarray}
r_1&=& -  l - M + \frac{3M^2-q^2}{2 l} +  \left(a^2 - 4 M^2+2 q^2 \right ) \frac{M}{l^2}\,, \\[2mm]
r_2 &=& {\tilde r}_{-} \,+ \, \frac{2 M {\tilde r}_{-}
-q^2}{{\tilde r}_{-} - M}\,\, \frac{{\tilde r}_{-}^{2}}{ 2 l^2}\,,
\\[2mm]
r_3 &=& {\tilde r}_{+}\, + \, \frac{2 M {\tilde r}_{+}
-q^2}{{\tilde r}_{+} - M}\,\, \frac{{\tilde r}_{+}^{2}}{ 2 l^2}\,,
\\[2mm]
r_4 &=& l - M  -\frac{3M^2 - q^2}{2 l} +\left(a^2 - 4 M^2+2 q^2 \right ) \frac{M}{l^2}\,,
\label{limits1}
\end{eqnarray}
where the radii
\begin{eqnarray}
\tilde{r}_{\pm}= M\pm \sqrt{M^2-a^2-q^2}
\label{hradii0}
\end{eqnarray}
correspond to the locations of the inner and outer horizons of the black hole in the limit $ l \to \infty $. We observe that the root $ r_1 $ is negative and must therefore be discarded as unphysical. Among the remaining roots, $r_2 = r_{-}$ describes the location of the black hole inner horizon, $r_3 = r_{+}$ corresponds to the outer (event) horizon
$( r_{-} < r_{+} < \tilde{r}_{+} )$, while $ r_4 $ gives the location of the cosmological horizon, $ r_4 = r_{++}  $. Clearly, in the limit $ l \to \infty $ the cosmological horizon is pushed arbitrarily far away from the black hole event horizon. We note that a similar analysis of the quartic horizon equation in the anti-de Sitter case $(\Lambda < 0)$, where only two physical roots exist, has been carried out in Ref. \cite{aliev1}.

We emphasize that the multi-horizon black hole configurations analyzed in this work are not regular (singularity-free) solutions. As in the Kerr-(Newman)-de Sitter spacetimes of general relativity, these geometries generically possess a curvature singularity at the center, independent of the presence of electric charge. The purpose of the present analysis is not to construct regular black holes, which typically require additional mechanisms such as nonlinear matter sources or modified electrodynamics  (see, for instance,  Ref. \cite{odin2}), but rather to investigate the analytic horizon structure and extremal properties of  Kerr-Newman-de Sitter black holes in $ f(R) $ gravity with constant curvature.

\subsubsection{Extremal  Configurations  I}  

Let us now consider extremal black holes corresponding to $M=M_{1e}$ and $M=M_{2e}$, respectively. It is straightforward to show that for $M=M_{1e}$,
the real root of the resolvent equation given in (\ref{resolvereal}) simplifies drastically and takes the form
\begin{eqnarray}
u &=& - \frac{l^2}{3} \left(1- \frac{a^2}{ l^2} -  \eta  \right).
\label{umerge1}
\end{eqnarray}
Using this expression in Eqs. (\ref{4rootsnot}) and (\ref{4roots}), we find that the outer and inner horizons coincide, resulting in an extremal configuration with a double root,
\begin{eqnarray}
 r_{-} &=&
r_{+} = \frac{l}{\sqrt{6}}
\left(1- \frac{a^2}{l^2} -\eta \right)^{1/2}\,.
\label{merge1}
\end{eqnarray}
In the limit $l \rightarrow \infty$, this expression reduces to the radius of the extremal horizon of the Kerr--Newman black hole, namely $r_{+}= \sqrt{a^2+q^2}$, as expected.

Similarly, for $M=M_{2e}$ we find that
\begin{eqnarray}
u &=& - \frac{l^2}{3} \left(1- \frac{a^2}{ l^2} -  2\eta  \right),
\label{umerge2}
\end{eqnarray}
which, upon substitution into the corresponding expressions   Eqs. (\ref{4rootsnot}) and (\ref{4roots}), leads to an extremal configuration with a single horizon. We obtain
\begin{eqnarray}
r_{+} &=&
r_{++}=\frac{l}{\sqrt{6}}
\left(1- \frac{a^2}{l^2} +\eta \right)^{1/2}.
\label{merge2}
\end{eqnarray}
That is, the black hole event horizon merges with the cosmological horizon. In the limit $l \rightarrow \infty$, this expression implies that the radius of the extremal cosmological horizon is given by
$r_{++}= l/\sqrt{3}$.
Using the defining relation (\ref{l}), this result can also be expressed in terms of the constant curvature, $r_{++}=2/\sqrt{R_0}$, or equivalently in terms of the cosmological constant, $r_{++}=1/\sqrt{\Lambda}$. We recall that the quantity $\eta$ appearing above is defined in Eq. (\ref{eta}).

It is worth noting that an alternative and complementary description of these extremal configurations can be obtained by simultaneously solving the equations
\begin{eqnarray}
\Delta_r &=& 0\,, \qquad
\frac{d\Delta_r}{dr}=0.
\label{simultextr}
\end{eqnarray}
These conditions ensure the merging of the relevant horizons, namely the inner and outer black hole horizons or the outer black hole and cosmological horizons. Indeed, using Eq. (\ref{quartic1}) and solving the above system with respect to the mass parameter $M$ and the horizon radius $r$, one straightforwardly recovers exactly the same extremal mass values and horizon locations as those given in Eq.~(\ref{exmasses}) and Eqs.~(\ref{merge1}),~(\ref{merge2}), respectively.

What is particularly remarkable, however, is that solving the system~(\ref{simultextr}) instead with respect to the squared rotation parameter $a^2$ and the inverse square of the curvature radius $l^{-2}$ leads to closed analytic expressions. This provides a compact and fully transparent characterization of the extremal configurations, which, to our knowledge, has not been presented in this form in the existing literature. In contrast to previous approaches, where the rotation parameter is typically expressed implicitly as a function of the cosmological constant and studied mostly numerically, our formulation yields explicit relations $ a^2(r) $ and $ l^{-2}(r) $. These expressions allow for a direct and systematic analytic exploration of the extremal parameter space and considerably simplify the analysis of its physical properties. We now turn to this description.
It is straightforward to show that the system of equations (\ref{simultextr}) admits the solutions
\begin{eqnarray}
\label{aaq}
a^2 &= & \frac{M r -2 r^2-q^2 \pm \sqrt{Z}}{2}\,,  
\\ [2mm]
l^{-2} & =& \frac{M r +2 r^2-q^2 \mp \sqrt{Z}}{2 r^4}\,,
\label{llq}
\end{eqnarray}
where 
\begin{equation}
Z = M r^2 \left(M+8 r \right) - q^2 \left(4 r^2 + 2 M r- q^2\right)
\label{Z}
\end{equation}
and the upper sign corresponds to the merger of the black hole inner and outer horizons, whereas the lower sign describes the merger of the outer horizon with the cosmological horizon. In what follows, we restrict our attention to the inner-outer horizon merger, which yields the physically relevant extremal black hole configurations discussed in this work.

Returning to Eqs. (\ref{aaq}) and (\ref{llq}), we observe that the extremal values of the rotation parameter $ a^2 $ and the inverse square of the curvature radius $l^{-2}$ depend explicitly on both the horizon location and the electric charge of the black hole. Moreover, only in the limit of vanishing background curvature, $l \to \infty$ (corresponding to the horizon location $ r=M $), does the extremality condition for the rotation parameter become universal, reducing to the familiar Kerr--Newman bound $ M^2 = a^2 + q^2 $. Away from this limit, the extremality conditions are intrinsically non-universal and encode the interplay between charge, rotation, and background curvature.
Nevertheless, one can still identify a distinguished \emph{ultra-extremal} configuration, defined as the configuration in which the squared rotation parameter attains its maximal value as a function of the horizon radius. This configuration is determined by the condition
\begin{eqnarray}
\frac{d\, a^2(r)}{dr} &=& 0 \,,
\label{aextrem}
\end{eqnarray}
whose solution yields the corresponding horizon location
\begin{eqnarray}
r &=& \frac{M}{2} \left( 1 + \sqrt{5}\,\sin\Phi \right) .
\label{r0max}
\end{eqnarray}
Here the dependence on the electric charge $q^2$ is entirely encoded in the function
\begin{eqnarray}
\Phi &=&   \frac{\pi}{6} +  \frac{1}{3} \, \arccos \left( \frac{8 q^2 - 11 M^2}{5 \sqrt{5} M^2}\right)\,.
\label{phi}
\end{eqnarray}
which will be specified explicitly below. 
For subsequent purposes, it is useful to expand Eq.~(\ref{r0max}) in powers of  $q^2 $. Restricting ourselves to the leading order, we obtain
\begin{eqnarray}
r &=& \frac{3 + 2\sqrt{3}}{4}\, M 
+ \left(\sqrt{3}-2\right)\frac{q^2}{3M} \, .
\label{expanr0max}
\end{eqnarray}
This expression shows that the contribution of the electric charge is negative, and therefore reduces the value of the horizon radius corresponding to the ultra-extremal configuration. In view of this result, it is then straightforward to infer from Eqs.~(\ref{aaq}) and (\ref{llq}) that the ultra-extremal value of the squared rotation parameter $a^2$ decreases monotonically as the electric charge increases, whereas the inverse square of the curvature radius $l^{-2}$ increases correspondingly with increasing charge. Indeed, substituting the expression (\ref{r0max}) into Eqs. (\ref{aaq}) and (\ref{llq}) and performing straightforward algebraic manipulations, these relations can be cast into the form
\begin{eqnarray}
\label{aphi} 
a^2 &=&   \frac{\left(1 +  \sqrt{5} \sin\Phi \right)\left( 7 M^2 - 8 q^2 +  5 \sqrt{5}\, M^2 \sin\Phi  - 10 M^2 \sin^2 \Phi \right)}{4 \left(1 + 2 \sqrt{5} \sin\Phi \right)}\,,
\\ [2mm]  
l^{-2} &=&    \,\frac{4 \left[4 q^2 +M^2 \left( 10 \sqrt{5}\, \sin^3\Phi   +5 \sin^2\Phi  - 6 \sqrt{5} \sin\Phi -5 \right) \right]}{M^4 \left(1 + \sqrt{5} \sin\Phi \right)^4  \left(1 + 2 \sqrt{5} \sin\Phi \right)}\,,
\label{lphi}
\end{eqnarray}
where $ \Phi $ is the same as defined in Eq. (\ref{phi}). Thus, we arrive at general analytic expressions that explicitly encode the dependence of the ultra-extremal value of the rotation parameter $a^2$ (and, correspondingly, of the inverse square of  the  curvature radius  $l^{-2}$) on the electric charge of the black hole.

Expanding these expressions in powers of $q^2$ and restricting ourselves to the lowest nontrivial order, we obtain
\begin{eqnarray}
\label{expanaphi}
a^2 &=&  \frac{3}{16} \left(3 + 2 \sqrt{3}\right) M^2 +\frac{1}{4} \left(\sqrt{3} -6 \right) q^2, \\ [2mm]  
l^{-2} &=&  \frac{16}{ \left(3 + 2 \sqrt{3}\right)^3 M^2} +  \frac{64 \left(362 -209\sqrt{3}\right)}{9 M^2}\, \frac{q^2}{M^2}\,.
\label{expanlphi}
\end{eqnarray}
From the first expression it follows that a nonvanishing electric charge contributes negatively to $a^2$, thereby reducing its value, whereas the second expression shows that the electric charge increases the value of $l^{-2}$.  This provides an explicit analytic confirmation of the general behavior discussed above.  

From Eq.~(\ref{aaq}) it also follows that at the horizon radius
\begin{eqnarray}
r &=& \frac{1}{2} \left( 3M + \sqrt{9 M^2 - 8 q^2} \right),
\label{rzeroa2}
\end{eqnarray}
the squared rotation parameter vanishes, $a^2=0$. This, in turn, yields the corresponding limiting value of the electric charge,
\begin{eqnarray}
q_{\rm max} &=& \sqrt{\frac{9}{8}}\, M .
\label{limitingq}
\end{eqnarray}
Using these values of $r$ and $q^2$ in Eq.~(\ref{llq}), we find
\begin{eqnarray}
l^{-2} &=& \frac{2}{27 M^2} = \frac{0.074074}{M^2} \,,
\label{limitingrn}
\end{eqnarray}
which can equivalently be expressed in terms of the scalar curvature 
(or the cosmological constant) as
\begin{eqnarray}
R_0 &=& \frac{8}{9 M^2}\,, \qquad
\Lambda = \frac{2}{9 M^2}\,.
\label{limiting2}
\end{eqnarray}
These quantities characterize the existence of nonrotating extremal charged black holes in $f(R)$ gravity and, equivalently, extremal Reissner–Nordström–de Sitter black holes in general relativity.  Having established these limiting configurations analytically, it is instructive to explore how the extremal values of $ a^2 $ and $ l^{-2}$ evolve as functions of the electric charge. The corresponding behavior, as determined by Eqs. (\ref{aphi}) and (\ref{lphi}), is illustrated in Figure 1. 
\begin{figure}[h!]
\centering
\begin{tabular}{cccc}
\epsfig{file=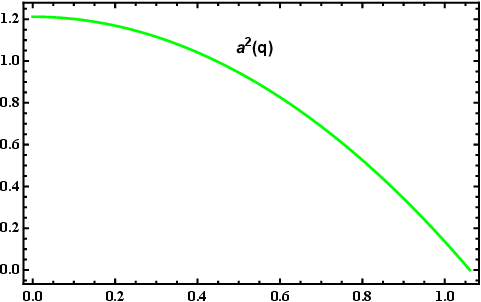,width=0.50\linewidth,clip=} &&&
\epsfig{file=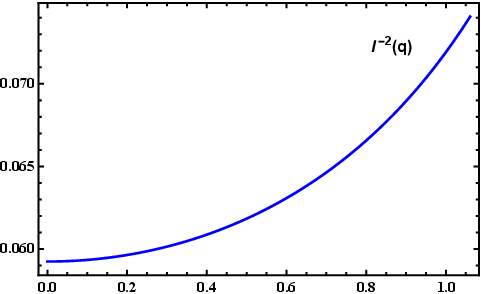,width=0.50\linewidth,clip=}
\end{tabular}
\caption{Dependence of the squared rotation parameter $a^2$ and the inverse square of the curvature radius $l^{-2}$ on the electric charge $q$ of the black hole. As the electric charge increases toward its limiting value, the quantity $a^2$ decreases monotonically, while $l^{-2}$ increases monotonically.}
\end{figure}%
We observe that as the electric charge increases from zero to its limiting value given in Eq. (\ref{limitingq}), the squared rotation parameter $a^2$ decreases monotonically from its ultra-extremal value  at zero charge  in Eq. (\ref{expanaphi}) to zero, whereas the inverse square of the curvature radius $l^{-2}$ increases monotonically from the value in Eq.~(\ref{expanlphi}) at zero charge  to the limiting value in Eq.~(\ref{limitingrn}).

To proceed further, it is useful to consider an illustrative example with electric charge $q = M/2 $. It is then straightforward to show that the inverse square of the curvature radius $l^{-2}$ vanishes at the horizon location $r = M$, while the familiar Kerr--Newman bound yields the maximal value  $a_{\max}^2 = 0.75\,M^2$. 
In this case, it also follows from Eq.~(\ref{rzeroa2}) that the squared rotation parameter vanishes, $a^2=0$, at the horizon location
\begin{eqnarray}
r &=& \frac{1}{2}\left( 3 + \sqrt{7} \right) M = 2.82288\,M .
\label{ru1}
\end{eqnarray}
Meanwhile, the inverse square of the curvature radius at this horizon location takes the value
\begin{eqnarray}
l^{-2} &=& \frac{14\sqrt{7} - 37}{M^2} = \frac{0.40518}{M^2} .
\label{aaqll0q3}
\end{eqnarray}
This, in turn, corresponds to a scalar curvature $R_0 = 0.48622/M^2$, or equivalently to a cosmological constant  $\Lambda = 0.12156/M^2$.

Next, solving the extremality condition (\ref{aextrem}) for $q = M/2$, it is straightforward to show that the squared rotation parameter $a^2$ attains its ultra-extremal value at the horizon location
\begin{eqnarray}
r &=& \frac{1}{8}\left( 7 + \sqrt{33} \right) M = 1.59307\,M .
\label{ruq}
\end{eqnarray}
We find
\begin{eqnarray}
a^2 &=& \frac{1}{32}\left( 13 + 3\sqrt{33} \right) M^2 
      = 0.94480\,M^2 \,, \qquad 
a = 0.97201\,M ,
\label{aaqll0q1}
\end{eqnarray}
while the corresponding value of the inverse square of the curvature radius is given by
\begin{eqnarray}
l^{-2} &=& \frac{259 - 45\sqrt{33}}{8M^2} 
        = \frac{0.06183}{M^2} .
\label{aaqll0q2}
\end{eqnarray}

It is important to emphasize that for horizon locations with  $r$ smaller than that given in Eq.~(\ref{ru1}), or equivalently for values of  $l^{-2}$ exceeding that in Eq.~(\ref{aaqll0q3}), the existence of a \emph{minimum rotation} of the black hole becomes unavoidable. In our approach, this feature  emerges naturally and admits a clear and unambiguous characterization. Specifically, the minimum rotation is determined by the intersection of the curves $a^2(r)$ and $l^{-2}(r)$, obtained from the analytic relations (\ref{aaq}) and (\ref{llq}). For definiteness, we determine this location by numerically solving the equation $a^2 - l^{-2} = 0$ for $M=1$ and $q=1/2$. The physical root of this equation corresponds precisely to the intersection point of the two curves. We find
\begin{eqnarray}
r = 2.79661\,M, \qquad
a^2 = 0.040963\,M^2 \; (a=0.202393\,M), \qquad
l^{-2} = \frac{0.040963}{M^2},
\label{physroot2}
\end{eqnarray}
or equivalently,
\begin{eqnarray}
R_0 = \frac{0.48622}{M^2}, \qquad
\Lambda = \frac{0.121555}{M^2}.
\label{physroot3}
\end{eqnarray}
The results of this analysis are displayed in Figure 2. For the electric charge
$q = M/2$, we recover the Kerr--Newman extremal value $a_{\max}^2 = 0.75$ together
with $l^{-2}=0$ at the horizon location $r=M$. As the horizon radius increases,
both $a^2$ and $l^{-2}$ grow monotonically and reach their ultra-extremal values,
indicated by the red points in the figure, given in Eqs.~(\ref{aaqll0q1}) and (\ref{aaqll0q2}) at the location (\ref{ruq}). For larger values of $r$, the quantity $a^2$ decreases and vanishes at the location given in Eq.~(\ref{ru1}). In parallel, $l^{-2}$ also decreases and intersects the $a^2$ curve at the point (\ref{physroot2}), marked by the orange dot. This intersection uniquely determines the minimum rotation of the black
hole, together with the corresponding scalar curvature or cosmological constant
given in Eqs.~(\ref{physroot2}) and (\ref{physroot3}).
\begin{figure}[h!]
\centering
\begin{tabular}{cccc}
\epsfig{file=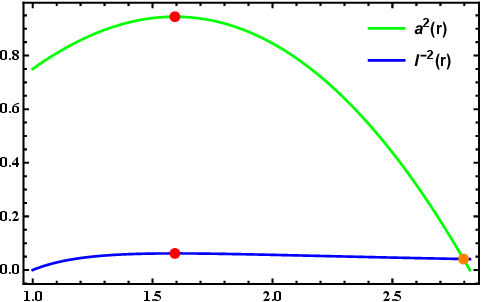,width=0.50\linewidth,clip=} &&&
\epsfig{file=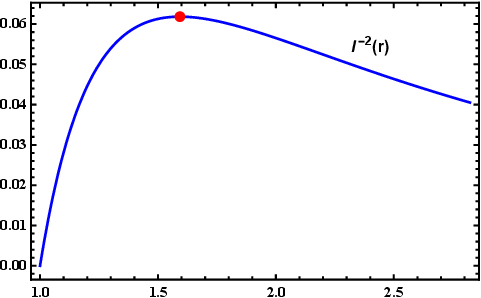,width=0.50\linewidth,clip=}
\end{tabular}
 \caption{Behavior of the squared rotation parameter $a^2$ and the inverse square
of the curvature radius $l^{-2}$ as functions of the horizon location $r$ for a
charged black hole with $q=M/2$. The red points indicate the ultra-extremal
values, while the orange point marks the intersection $a^2=l^{-2}$,
corresponding to the minimal rotation of the black hole. The right panel shows
$l^{-2}$ separately for improved visibility.} %
\end{figure}%
Thus, for charged black holes with $q=M/2$, the background curvature enforces a nonzero lower bound on the rotation, implying the existence of a minimal rotation parameter  $a_{\min}=0.202393,M$, realized precisely at the critical value of $l^{-2}$ (or, equivalently, $R_0$ and $\Lambda$).

Before concluding this subsection, it is also instructive to consider the  $q=0$ limit of the above analysis. In this case, Eqs.~(\ref{aaq}) and (\ref{llq}) reduce to particularly simple expressions,
\begin{eqnarray}
\label{aa0}
a^2 &=& \frac{1}{2}\, r \left( M - 2 r \pm \sqrt{M}\sqrt{M + 8 r} \right),  
\\[2mm]
l^{-2} &=& \frac{M + 2 r \mp \sqrt{M}\sqrt{M + 8 r}}{2 r^3}\,.
\label{ll0}
\end{eqnarray}
As in the case of nonvanishing electric charge, we observe that the extremal value 
of the squared rotation parameter $a^2$ explicitly depends on the horizon location 
$r$, reflecting the influence of the nonzero background curvature $R_0$ (or, 
equivalently, the cosmological constant $\Lambda$). Consequently, the familiar Kerr 
bound $a_{\rm max}=M$ is recovered only in the limit of vanishing background curvature,  corresponding to $l\to\infty$ and $r=M$.  Nevertheless, the analytic structure of Eqs.~(\ref{aa0}) and (\ref{ll0}) again allows  for the identification of an ultra-extremal configuration. In this case, the squared  rotation parameter $a^2$ attains a maximal value at a finite horizon location and then decreases monotonically to zero as the horizon radius increases. This behavior closely  parallels that found in the charged case and further illustrates the utility of our analytic approach in characterizing extremal and ultra-extremal black hole  configurations.

Next, from Eq.~(\ref{expanr0max}) in the $q=0$ limit, we find that the horizon location
\begin{eqnarray}
r &=& \frac{3 + 2 \sqrt{3}}{4}\, M = 1.61603\, M
\label{ru}
\end{eqnarray}
corresponds to the ultra-extremal configuration, at which both the squared rotation
parameter $a^2$ and the inverse square of the curvature radius $l^{-2}$ attain their
maximal values. Using Eqs.~(\ref{expanaphi}) and (\ref{expanlphi}), we obtain
\begin{eqnarray}
\label{aa0ll01}
a^2 &=& \frac{3}{16} \left(3 + 2 \sqrt{3}\right) M^2
       = 1.21202\, M^2 \qquad (a = 1.10092\, M),
\\[4mm]
l^{-2} &=& \frac{16}{\left(3 + 2 \sqrt{3}\right)^3 M^2}
          = \frac{0.0592373}{M^2}\, .
\label{aa0ll02}
\end{eqnarray}
Equivalently, these values correspond to the scalar curvature and cosmological constant
\begin{eqnarray}
R_0 = \frac{0.710848}{M^2}\,, \qquad
\Lambda = \frac{0.177712}{M^2}\,.
\label{rlam}
\end{eqnarray}

Meanwhile, the $q=0$ limit of Eq.~(\ref{rzeroa2}) shows that the squared rotation
parameter vanishes, $a^2=0$, at the horizon location $r=3M$. Substituting this value
into Eq.~(\ref{ll0}), we obtain
\begin{eqnarray}
l^{-2} &=& \frac{1}{27 M^2}\,,
\label{limiting1}
\end{eqnarray}
which corresponds to the following values of the scalar curvature and the
cosmological constant,
\begin{eqnarray}
R_0 &=& \frac{4}{9 M^2}\,, \qquad
\Lambda = \frac{1}{9 M^2}\,.
\label{limiting02}
\end{eqnarray}
These quantities characterize the limiting values for the existence of nonrotating black holes in $f(R)$ gravity with constant curvature, or equivalently Schwarzschild--de Sitter black holes in GR. Thus, even in the absence of electric charge, the presence of a positive
background curvature enforces a finite range of admissible rotations, bounded from above by an ultra-extremal configuration and from below by the nonrotating Schwarzschild-de Sitter limit. We note that the results given in Eq.~(\ref{ru}) and Eqs.~(\ref{aa0ll01})–(\ref{aa0ll02}) are consistent with those obtained, in a slightly different setting, in \cite{slany} for the Kerr–de Sitter spacetime.

Finally, we note that the existence of a minimal rotation is unavoidable even
in the absence of electric charge. Specifically, for horizon locations with
$r<3M$, or equivalently for values of the background curvature $R_0$ (or
$\Lambda$) exceeding those given in Eq.~(\ref{limiting02}), a nonzero lower
bound on the rotation parameter necessarily emerges. In close analogy with the charged case, this minimal rotation can be identified as the intersection point of the curves $a^2(r)$ and $l^{-2}(r)$, constructed from the analytic expressions (\ref{aa0}) and (\ref{ll0}). The intersection is determined by numerically solving the equation $a^2-l^{-2}=0$, where we set $M=1$. The physically relevant root is found to be
\begin{eqnarray}
 r = 2.97908 M, \qquad a^2 = 0.0373485 M^2 \; (a = 0.193258 M), \qquad l^{-2} = \frac{0.0373485}{M^2}\,. 
 \label{physroot1} 
 \end{eqnarray}
This provides a concrete and geometrically transparent definition of minimal
rotation, arising naturally from the analytic structure of the extremality conditions.
Thus, for this critical value of the horizon radius $ r $ and the corresponding value of
$ l^{-2} $ (or equivalently, $ R_0 = 0.448182/M^2 $ and $ \Lambda = 0.112046/M^2 $),
black holes in such a universe are necessarily rotating, with a minimum rotation
parameter $ a_{\min}=0.193258\,M $. Comparing these results with those obtained above for the charged case $ q=M/2 $ (see Eqs.~(\ref{physroot2})--(\ref{physroot3})), we find that the presence of electric charge increases the minimum rotation required for the existence of the black hole. The results of this analysis are displayed in Figure 3.
\begin{figure}[h!]
\centering
\begin{tabular}{cccc}
\epsfig{file=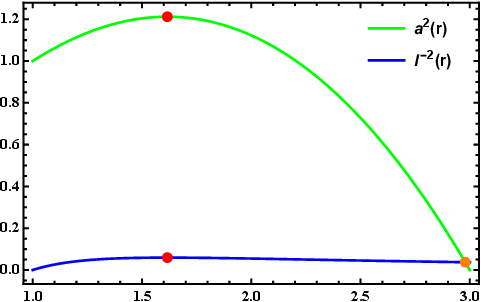,width=0.50\linewidth,clip=} &&&
\epsfig{file=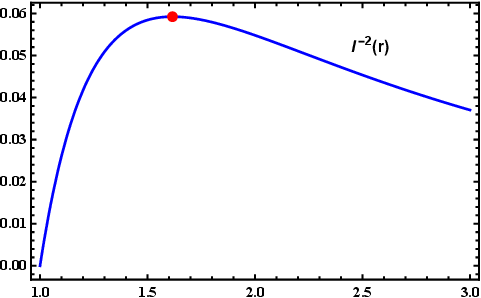,width=0.50\linewidth,clip=}
\end{tabular}
\caption{Dependence of the squared rotation parameter $ a^2 $ and the inverse square of the curvature radius $ l^{-2} $ on the horizon location $ r $ of the black hole for vanishing electric charge. The left panel shows the simultaneous behavior of both quantities, while the right panel highlights the variation of $l^{-2}$ for clarity.}%
\end{figure}%
We observe that as the horizon location increases from $r=M$ to $r=3M$, the squared rotation parameter $a^2$ initially grows monotonically from its universal general relativistic value $a^2=M^2$. It reaches its ultra-extremal value, given in Eq.~(\ref{aa0ll01}), at the location (\ref{ru}), after which it decreases and eventually vanishes at $r=3M$. Meanwhile, the inverse square of the curvature radius $l^{-2}$ which is proportional to the scalar curvature (or, equivalently, the cosmological constant) increases monotonically from zero and attains its maximum at the same horizon location where $a^2$ reaches its ultra-extremal value (indicated by the red points on the curves). Beyond this point, $l^{-2}$ decreases as $r$ approaches $3M$, intersecting the $a^2$ curve at the location given in Eq.~(\ref{physroot1}), marked by the orange point. For clarity, the variation of $l^{-2}$ over the interval $M \le r \le 3M$ is shown separately in the right panel, where the red point denotes its maximal value. In the left panel, the simultaneous presence of two curves with different scales renders this maximum less clearly visible.

We note that the above results are in qualitative agreement with those obtained in \cite{akcay, cembra}, where the existence of a minimal rotation was inferred from numerical analyses of the cumbersome functions $ a(R_0) $ and $ a(\Lambda)$, respectively.

\subsection{Horizons II}

Returning to the resolvent cubic equation~(\ref{cubic1}), it is straightforward to
show that the corresponding Vieta relations take the form
\begin{eqnarray}
u_1 + u_2 + u_3 &=& a^2 - l^2 \,, \nonumber \\[2mm]
u_1 u_2 + u_1 u_3 + u_2 u_3 &=& 4 l^2 (a^2 + q^2) \,, \nonumber \\[2mm]
u_1 u_2 u_3 &=& 4 l^4 \left[ M^2 - (a^2 + q^2)\left(1 - \frac{a^2}{l^2}\right) \right].
\label{rvieta}
\end{eqnarray}
From the last relation in~(\ref{rvieta}) it follows that, when the condition
\begin{equation}
M^2 = (a^2 + q^2)\left(1 - \frac{a^2}{l^2}\right)
\label{constraint}
\end{equation}
is satisfied, the resolvent cubic equation admits a vanishing root, $u=0$.
This shows that the above mass constraint arises naturally from the algebraic
structure of the resolvent equation itself, rather than being imposed by hand. Substituting $u=0$ into the remaining Vieta relations, one readily verifies that
the two remaining roots are determined by a quadratic equation. In what follows,
we adopt the root $u=0$ for definiteness, noting that choosing any of the other
roots leads to an equivalent description and does not alter the qualitative features of the horizon structure. Remarkably, the same condition~(\ref{constraint}) also leads to a drastic simplification of the quartic horizon equation~(\ref{quartic1}). When imposed,
the quartic polynomial becomes factorizable, revealing a highly constrained root structure that, as we shall show, allows only a specific type of horizon merger in the extremal regime. Explicitly, one finds
\begin{equation}
 r^4   - l^2 \left( \sqrt{1-\frac{a^2}{l^2}}\,\,   r     - \sqrt{a^2+q^2} \right)^2 = 0\,.
 \label{quartic2}
\end{equation}
Consequently, using either the general expressions (\ref{4rootsnot}) and (\ref{4roots})  with $u=0$, or equivalently the factorization of the quartic equation~(\ref{quartic2}) into two quadratic polynomials, we obtain two independent pairs of roots of the quartic equation~(\ref{quartic1}). The first pair of roots is given by
\begin{equation}
r_{1,2} = -\frac{l}{2}
\left(
\sqrt{1-\frac{a^2}{l^2}}
\pm
\sqrt{1-\frac{a^2}{l^2}
+ \frac{4}{l}\,\sqrt{a^2+q^2}}
\right),
\label{roots12}
\end{equation}
where the upper sign corresponds to the negative root $r=r_1$, which is
unphysical and must therefore be discarded.

The second pair of roots takes the form
\begin{equation}
r_{3,4} =
\frac{l}{2}
\left(
\sqrt{1-\frac{a^2}{l^2}}
\mp
\sqrt{1-\frac{a^2}{l^2}
- \frac{4}{l}\,\sqrt{a^2+q^2}}
\right),
\label{roots34}
\end{equation}
and represents the physically relevant horizons. The physical interpretation of these roots becomes more transparent by considering their expansion in powers of the curvature radius $l$ in the limit $l \rightarrow \infty$. Restricting ourselves to terms up to second
order in $l^{-1}$, we find from Eqs.~(\ref{roots12}) and (\ref{roots34}), respectively, that
\begin{eqnarray}
r_2=  r_{-} =  \sqrt{a^2+q^2} - \frac{a^2+q^2 }{l}  + \frac{(5 a^2+ 4 q^2)\sqrt{a^2+q^2}}{2 l^2}\,, 
\label{expansions0}\\[3mm]
r_3= r_{+} =  \sqrt{a^2+q^2}  + \frac{a^2+q^2 }{l}  + \frac{(5 a^2+ 4 q^2)\sqrt{a^2+q^2}}{2 l^2}\,,
\label{expansions00} \\[3mm]
 r_4=r_{++} = l - \sqrt{a^2+q^2} - \frac{3 a^2 +2 q^2} {2 l} - \frac{(5 a^2+ 4 q^2) \sqrt{a^2+q^2}}{2 l^2} .
 \label{expansions2}
\end{eqnarray}
Comparing these expressions with those in Eqs.~(\ref{roots12}) and (\ref{roots34}), we conclude that the lower sign in Eq.~(\ref{roots12}) corresponds to the location $r_{-}$ of the inner horizon. In contrast, in Eq.~(\ref{roots34}) the root with the upper sign describes the position $r_{+}$ of the outer black hole horizon, while the root with the lower sign gives the location $r_{++}$ of the cosmological horizon.  In the limit of vanishing scalar curvature $R_0$, or equivalently vanishing cosmological constant, i.e.\ $l \rightarrow \infty$, it follows from Eqs.~(\ref{expansions0}) and (\ref{expansions00}) that the inner and outer black hole horizons merge according to $r_{-} \rightarrow r_{+} = M$. This behavior is fully consistent with the mass constraint (\ref{constraint}) in this limit. In other words, the expansions in Eqs.~(\ref{expansions0})–(\ref{expansions2}) are performed around the extremal Kerr–Newman configuration, characterized by
$M^2 = a^2 + q^2$.

\subsubsection{Extremal Configurations II}

As shown above, the mass constraint (\ref{constraint}) effectively splits the roots of the quartic horizon equation into two disjoint pairs. As a direct consequence, only a single type of extremal rotation is permitted. Indeed, one readily verifies that the vanishing of the discriminant associated with Eq.~(\ref{roots34}) yields
\begin{equation}
1-\frac{a^2}{l^2} = \frac{4}{l}\, \sqrt{a^2+q^2},
\label{extreme34}
\end{equation}
whereas the discriminant corresponding to Eq.~(\ref{roots12}) never vanishes
for physical values of the parameters.  This condition gives rise to a degenerate extremal configuration characterized by a single horizon, resulting from the merger of the outer black hole horizon with the cosmological horizon. The coincident horizon is located at
\begin{equation}
r_{+}=r_{++} = \frac{l}{2}\, \sqrt{1-\frac{a^2}{l^2}}\,.
\label{extremeh}
\end{equation}
In the limit $ l\rightarrow \infty $, Eq.~(\ref{extremeh}) implies that the position of the extremal cosmological horizon approaches $ r_{++}= l/2 $. Using the relation in (\ref{l}), this result can be expressed in terms of the constant scalar curvature $ R_0 $ or, equivalently, the cosmological constant $\Lambda$, yielding
\begin{equation}
r_{++}= \sqrt{\frac{3}{R_0}}
\qquad \text{or} \qquad
r_{++}= \frac{1}{2}\,\sqrt{\frac{3}{\Lambda}} \, .
\label{rrlambda}
\end{equation}
Moreover, it is clear that under the extremality condition (\ref{extreme34}), the mass determined by the constraint (\ref{constraint}) also becomes extremal and must coincide with the upper extremal mass $ M_{2e} $ introduced in Eq.~(\ref{exmasses}). Indeed,
one readily verifies that
\begin{equation}
M^2 = M_{2e}^2
= \frac{l^2}{16}\left(1-\frac{a^2}{l^2}\right)^3 .
\label{extreme3}
\end{equation}
In the limit $ l\rightarrow \infty $, this expression diverges as $ M_{e}^2 \rightarrow l^2/16 $. This behavior is to be contrasted with the result $ M_{2e}^2 \rightarrow l^2/27 $, which arises in the absence of the mass constraint (\ref{constraint}) (see Eq.~(\ref{exmasses})).

We note that the above expressions may be rewritten in alternative but equivalent forms in terms of the rotation parameter $a$ and the electric charge $q$. In particular, solving Eq.~(\ref{extreme34}) for the curvature radius yields
\begin{equation}
l^{-1} =
\frac{\sqrt{5a^2+4q^2}-2\sqrt{a^2+q^2}}{a^2}\,.
\label{laq}
\end{equation}
Substituting this expression into Eqs.~(\ref{extremeh}) and (\ref{extreme3}), we obtain the compact relations
\begin{equation}
r_{+}=r_{++}
= \sqrt{\,2(a^2+q^2)
+ \sqrt{a^2+q^2}\,\sqrt{5a^2+4q^2}} \,,
\label{extremeh1}
\end{equation}
and
\begin{equation}
M^2 = M_{2e}^2
= \frac{4 (a^2+q^2)^{3/2}}
{2\sqrt{a^2+q^2}+\sqrt{5a^2+4q^2}} \, .
\label{extreme4}
\end{equation}
It is worth emphasizing that the same results can be obtained independently by simultaneously solving the system of equations (\ref{simultextr}). This alternative derivation confirms the internal consistency of the analysis and provides a complementary perspective on the extremal horizon structure.

Thus, within the constrained sector (\ref{constraint}), the theory admits a unique extremal configuration, excluding the possibility of an inner–outer horizon merger. In this sense, the mass constraint induces a chiral-like structure in the horizon configuration space: only one type of horizon merger, namely the outer–cosmological one, is permitted, while the inner–outer merger is excluded.

\section{Conclusion}

In this paper, we have presented a unified analytic investigation of the horizon structure and extremal properties of rotating and charged black holes in $f(R)$ gravity with constant curvature, focusing on the Kerr--Newman--de Sitter class of solutions. Using the mapping between constant curvature solutions of general relativity and their counterparts in $f(R)$ gravity, we were able to treat both frameworks systematically within a single analytic scheme and to provide a transparent comparison between them.

We have solved the quartic horizon equation analytically and derived closed-form expressions for the horizon radii, clarifying their physical interpretation in terms of inner, outer, and cosmological horizons. Our analysis shows that, in the presence of a nonvanishing background curvature, the extremality conditions are generically non-universal and depend explicitly on the horizon location and the electric charge. The familiar Kerr--Newman bound is recovered only in the limit of vanishing background curvature.

A central result of this work is the derivation of explicit analytic expressions for the squared rotation parameter $a^2$ and the inverse square of the curvature radius $l^{-2}$ as functions of the horizon location and the electric charge. This formulation provides a clear and efficient framework for studying extremal configurations, allowing us to identify ultra-extremal solutions in which $a^2$ attains its maximal value and subsequently decreases as the charge or the horizon radius is varied. In contrast to earlier approaches based on cumbersome functions $a(\Lambda)$ or $a(R_0)$ and basically numerical analyses, our method yields compact analytic relations with direct physical interpretation.

An important consequence of our analysis is the existence of a minimum rotation parameter for black holes embedded in a universe with sufficiently large positive curvature. For both charged and uncharged cases, we showed that this minimum rotation arises naturally as the intersection point of the curves $a^2(r)$ and $l^{-2}(r)$. In particular, for charged black holes with $q = M/2$, as well as in the neutral limit $q = 0$, the presence of a positive background curvature enforces a strictly nonzero lower bound on the rotation, a feature that is absent in asymptotically flat spacetimes.

Finally, we have demonstrated that imposing a specific mass constraint leads to a factorization of the quartic horizon equation and induces a chiral-like structure in the space of horizon configurations. In this constrained sector, only one type of extremal configuration is allowed, corresponding to the merger of the outer black hole horizon with the cosmological horizon, while the inner-outer horizon merger is excluded. This provides a novel and physically transparent characterization of extremality in rotating de Sitter black hole spacetimes.

Our analysis can be extended in several directions. First, it would be interesting to investigate ergoregions in the spacetime of Kerr-Newman-de Sitter black holes using the unified analytic treatment developed here. Second, it would also be of interest to extend our framework to Kerr–Newman–anti–de Sitter black holes in general relativity and in $f(R)$ gravity with constant negative curvature. In addition, extending our analysis to geodesic motion and the associated quasi-periodic oscillations in the spacetime of Kerr-Newman-de Sitter black holes would be valuable for the study of potential observational signatures of these systems, particularly in extremal regimes.


\end{document}